\RequirePackage{fix-cm}
\documentclass[smallextended]{svjour3}       
\usepackage[italian,english]{babel}
\smartqed  
\usepackage{hyperref}
\usepackage{xspace}
\usepackage{amssymb}
\usepackage{amsmath}
\usepackage{lipsum}
\usepackage{dsfont}
\usepackage{multirow}
\usepackage{afterpage}
\usepackage{amsfonts}
\usepackage{bm}%
\usepackage{babel}
\usepackage{graphicx}
\usepackage{color}

 \usepackage{mathptmx}      
\usepackage{latexsym}
\begin{document}

\title{Robustness of Measurement--Induced Correlations Under Decoherence Effect 
}


\author{S. Bhuvaneswari \and R. Muthuganesan        \and
       R. Radha 
}


\institute{Centre for Nonlinear Science (CeNSc),  Government College for Women, Kumbakonam - 612 001, Tamil Nadu, India \at
              \email{bhuvanajkm85@gmail.com}           
            \and
         Centre for Nonlinear Science \& Engineering, School of Electrical \& Electronics Engineering,
SASTRA Deemed University, Thanjavur, Tamil Nadu 613 401, India\at
              \email{rajendramuthu@gmail.com}
              \and
           Centre for Nonlinear Science (CeNSc),  Government College for Women, Kumbakonam - 612 001, Tamil Nadu, India \at
           \email{vittal.cnls@gmail.com}
}

\date{Received: date / Accepted: date}

\maketitle

\begin{abstract}

In this article, we study the dynamics of quantum correlation measures such as entanglement and measurement--induced nonlocality (MIN). Starting from an arbitrary Bell-diagonal mixed states under Markovian local noise such as bit-phase flip, depolarizing and generalized amplitude damping channel, we provide the decays of the entanglement measured by concurrence and quantum correlation captured by different forms of MIN (trace distance, Hilbert--Schmidt norm and relative entropy) as a function of the decoherence parameters. The effect of local noises on the dynamical behaviors of quantum correlation is observed. We show the existence of specific and important features of MIN such as revival, noise robustness and sudden change with respect to decoherence parameter. It is observed that all the noises cause sudden death of entanglement for partially entangled states. Further, we show the existence of separable quantum states with non-zero quantum correlations in terms of MIN.

\end{abstract}

\section{Introduction}
\label{intro}
Nonlocality, a peculiar property of quantum system plays an important role in understanding the characteristics of composite quantum systems and makes a fundamental departure from classical regime. It is a primary resource for various quantum information processing. In the framework of resource theory of quantum information, different resources have been introduced and explored in detail. A few notable resources are coherence \cite{Baumgratz,coherence}, entanglement \cite{Einstein1935,Schrodinger1935,Bell1964}, quantum discord \cite{Ollivier2001}, geometric discord \cite{Dakic2010}, measurement-induced nonlocality (MIN) \cite{SLUO2011}, etc.  Entanglement is the most peculiar property of quantum systems and is proven to be a vital ingredient for many quantum information applications including teleportation, cryptography and quantum communication \cite{Bennett,Bouwmeester,Bennett1992,Ekert,Shor}. Originally, it is believed that the entanglement is an important manifestation of nonlocal aspects of quantum systems and cannot be explained using local hidden variable theory \cite{Bell1964}. The detailed investigation and seminal work of Werner \cite{Werner} on this course reveals that entanglement cannot capture all the quantumness of correlations. It can only capture special kind of  correlations of composite quantum systems. The simplest example is that a separable state do not have entanglement between its constituents but possesses other forms of quantum correlations (beyond entanglement), which is measured through quantum discord \cite{Ollivier2001} and MIN \cite{SLUO2011}. This measure is shown to have advantage over the entanglement in deterministic quantum computational models using highly mixed  separable states and one qubit state.  It is known that quantum discord is not calculable in closed form, even for an arbitrary two-qubit state \cite{Girolami}.

Luo and Fu \cite{SLUO2011} identified a new variant of geometric measure of quantum correlation for bipartite systems due to locally invariant von Neumann projective measurements  as measurement induced nonlocality (MIN) \cite{SLUO2011}. This quantity captures the maximal nonlocal or global effects of the quantum system. In other words, this quantity is measured in terms of the maximal distance between the pre-- and post--measurement states using Hilbert--Schmidt distance. This measure is in some sense dual to the geometric measure of quantum discord. Hence, the notion of nonlocality may also be considered to be more general than that of Bell's version of nonlocality. Due to local ancilla problem \cite{Piani2012}, this measure is not considered as a bonafide candidate of a quantum correlation measure. In recent times, a different version of MIN has been proposed such as relative entropy \cite{ZXi}, fidelity \cite{Muthu1}, skew information \cite{Li}  and trace distance \cite{Hu2015} to counter the local ancilla problem. 

In general, quantum systems are always coupled with the environment and due to this unavoidable coupling, the system loses its unique signatures. In resource theory, these environments are modelled as noise or decoherence \cite{Nielsen2010}. The quantum systems are highly susceptible to noise. To perform any 
quantum information processing, it requires quantum resources which  will be robust against noise or decoherence. If we consider the entanglement as a resource for quantum information processing, while coupling with the environments, the entanglement between the constituents of the composite system decay with the noise or decoherence parameter. Under peculiar circumstances, the entanglement completely vanishes and is known as sudden death of entanglement \cite{Eberly2009}. In the light of the above observation, the researchers claim that entanglement is not found to be quantum advantageous.

In this paper, we study the dynamical behaviors of quantum correlations captured by entanglement and measurement-induced nonlocality (based on relative entropy and trace distance) under generalized amplitude damping and depolarizing noisy channels. It is observed that the both the noise cause sudden death  of entanglement. The observation reveals that the entropic and trace distance MIN are more robust than the entanglement under decoherence. 
\section{Quantum correlation measures}

\label{correlation}

\textit{Entanglement}

Let us consider a composite system $\rho$ shared by the subsystem $a$ and $b$ in the Hilbert space $\mathcal{H}=\mathcal{H}^a\otimes\mathcal{H}^b$. The amount of entanglement associated with a given two-qubit state $\rho$ can be quantified using concurrence \cite{Hill1997}, which is defined as 
\begin{equation}
C(\rho )=\text{max}\{0,~\lambda_1-\lambda_2-\lambda_3-\lambda_4\}
\end{equation}
where $\lambda_i$ are square root of eigenvalues of matrix $R=\rho \tilde{\rho }$ arranged in decreasing order. Here $\tilde{\rho } $ is spin flipped density matrix, which is defined as $\tilde{\rho }=(\sigma _y \otimes \sigma _y)\rho^{*}(\sigma _y \otimes \sigma _y)$. The symbol $*$ denotes the usual complex conjugate in computational basis. It is known that the concurrence varies from $0$ to $1$ with minimum and maximum values corresponding to separable and maximally entangled states respectively.
 
\textit{Measurement induced nonlocality}

A new measure is introduced to capture all the nonlocal effects that can be induced by local measurements, namely  measurement-induced nonlocality in terms of the Hilbert-Schmidt norm. It is originally defined as maximal square of Hilbert--Schmidt norm of difference of pre-- and post-- measurement states. It is defined as \cite{SLUO2011}
\begin{equation}
 N(\rho ) =~^{\text{max}}_{\Pi ^{a}}\| \rho - \Pi ^{a}(\rho )\| ^{2} 
\end{equation}
where the maximum is taken over the von Neumann projective measurements on subsystem $a$. Here $\Pi^{a}(\rho) = \sum _{k} (\Pi ^{a}_{k} \otimes   \mathds{1} ^{b}) \rho (\Pi ^{a}_{k} \otimes    \mathds{1}^{b} )$, with $\Pi ^{a}= \{\Pi ^{a}_{k}\}= \{|k\rangle \langle k|\}$ being the projective measurements on the subsystem $a$, which does not change the marginal state $\rho^{a}$ locally i.e., $\Pi ^{a}(\rho^{a})=\rho ^{a}$. If $\rho^{a}$ is non-degenerate, then the maximization is not required. 

Measurement induced nonlocality has recently attracted the attention of researchers due to easy computation and also its realization. However, it can change arbitrarily and reversibly through actions of the unmeasured party --local ancilla problem \cite{Piani2012}. Due to local ancilla problem, MIN is not a bonafide measure of quantum correlation. To address this issue, different forms of MIN have been identified using relative entropy, trace distance, fidelity and affinity.  

\textit{Relative entropy based MIN}

A natural way to circumvent this local ancilla problem is to define the correlation measure in terms of relative entropy and is defined as \cite{ZXi}
\begin{equation}
N_{RE}(\rho)= ~^{\text{max}}_{\Pi ^{a}} S\left( \rho||\Pi ^{a}(\rho )\right)
\end{equation}
where the maximum is taken over all the von Neumann measurements. Here, $S(X||Y)=\text{TrX}(\text{log}X-\text{log}Y)$ is relative entropy between $X$ and $Y$, $S(X)=-\text{Tr}X\text{log}Y$ is the von Neumann entropy, $\text{Tr}$ denotes trace of matrix and the logarithm is base 2. The quantity $N_{RE}(\rho)$ satisfies all the necessary requirements of a quantum correlation measure. Using the property $S( \rho\otimes P_{\alpha}||\sigma \otimes P_{\alpha})=S( \rho||\sigma)$, one can show that the relative entropy based MIN fixes the local ancilla problem.  Hence, the relative entropy based MIN is a bonafide measure of quantum correlation measure. Operationally, this quantity is interpreted as maximal entropy which increases due to local von Neumann projective measurement. 

\textit{Trace distance based MIN}

Other alternate form of MIN is based on trace distance \cite{Hu2015},  namely trace MIN (T-MIN)  which resolves the local ancila problem  \cite{Piani2012}. It is defined as
\begin{equation}
N_1(\rho):= ~^{\text{max}}_{\Pi^a}\Vert\rho-\Pi^a(\rho)\Vert_1
\end{equation} 
where $\Vert A \Vert_1 = \text{Tr}\sqrt{A^{\dagger}A}$ is the trace norm of operator $A$. Here also, the maximum is taken over all von Neumann projective measurements. For any $2\otimes2$ dimensional system, the closed formula of trace MIN $N_1(\rho)$ is given as 
\begin{equation}
N_1(\rho)=
\begin{cases}
\frac{\sqrt{\chi_+}~+~\sqrt{\chi_-}}{2 \Vert \textbf{x} \Vert_1} & 
 \text{if} \quad \textbf{x}\neq 0,\\
\text{max} \lbrace \vert c_1\vert,\vert c_2\vert,\vert c_3\vert\rbrace &  \text{if} \quad \textbf{x}=0,
\end{cases}
\end{equation}
where $\chi_\pm~=~ \alpha \pm 2 \sqrt{\tilde{\beta}} \Vert \textbf{x} \Vert_1 ,\alpha =\Vert \textbf{c} \Vert^2_1 ~\Vert \textbf{x} \Vert^2_1-\sum_i c^2_i x^2_i,\tilde{\beta}=\sum_{\langle ijk \rangle} x^2_ic^2_jc^2_k, \vert c_i \vert $ are the absolute values of $c_i$ and the summation runs over cyclic permutation of $\lbrace 1,2,3 \rbrace$ and $\text{x}_i$ are the components of vector $\textbf{x}$.

\section{Dynamics}
\begin{figure*}[!ht]
\centering\includegraphics[width=0.6\linewidth]{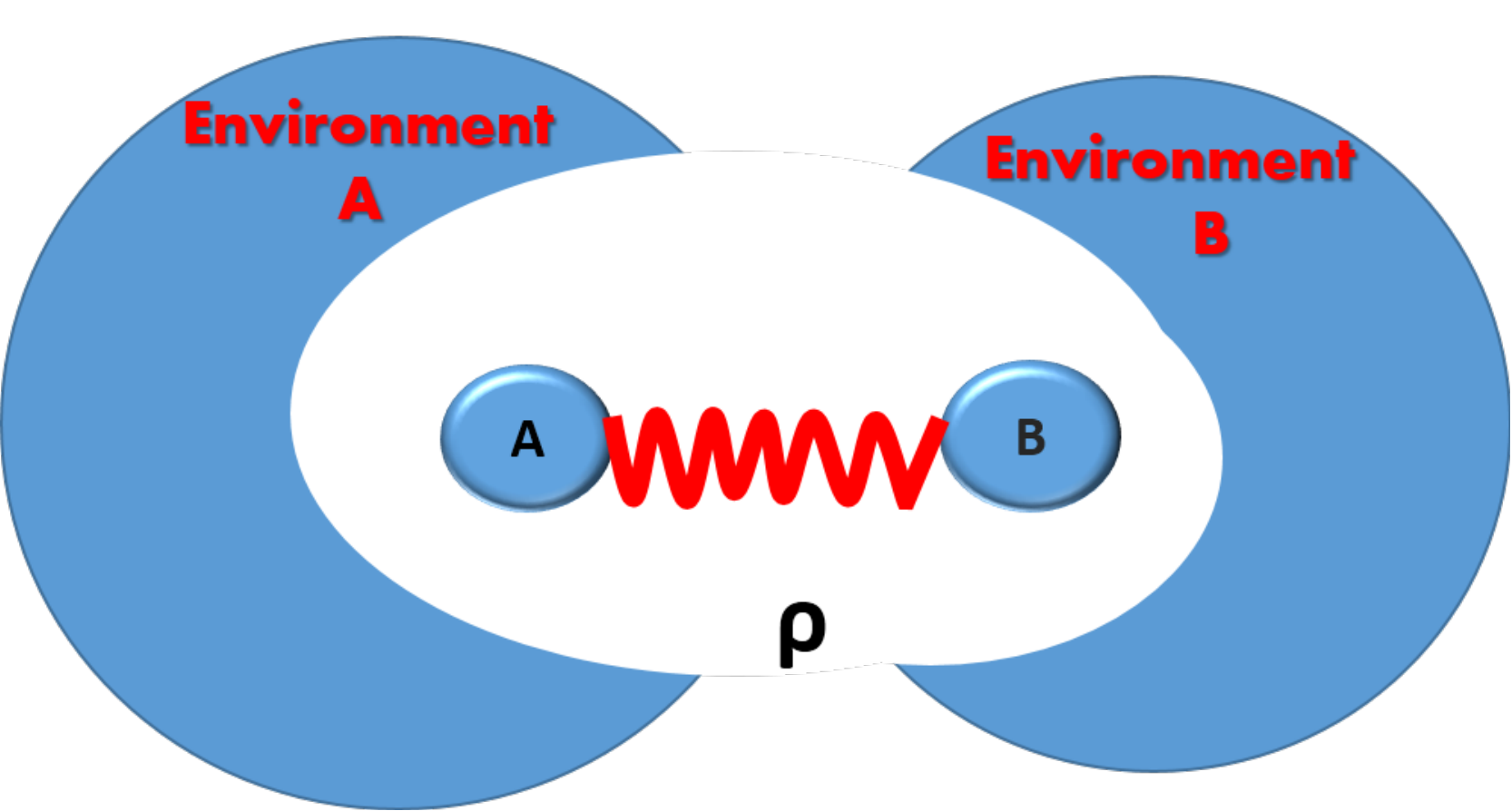}
\centering\includegraphics[width=0.7\linewidth]{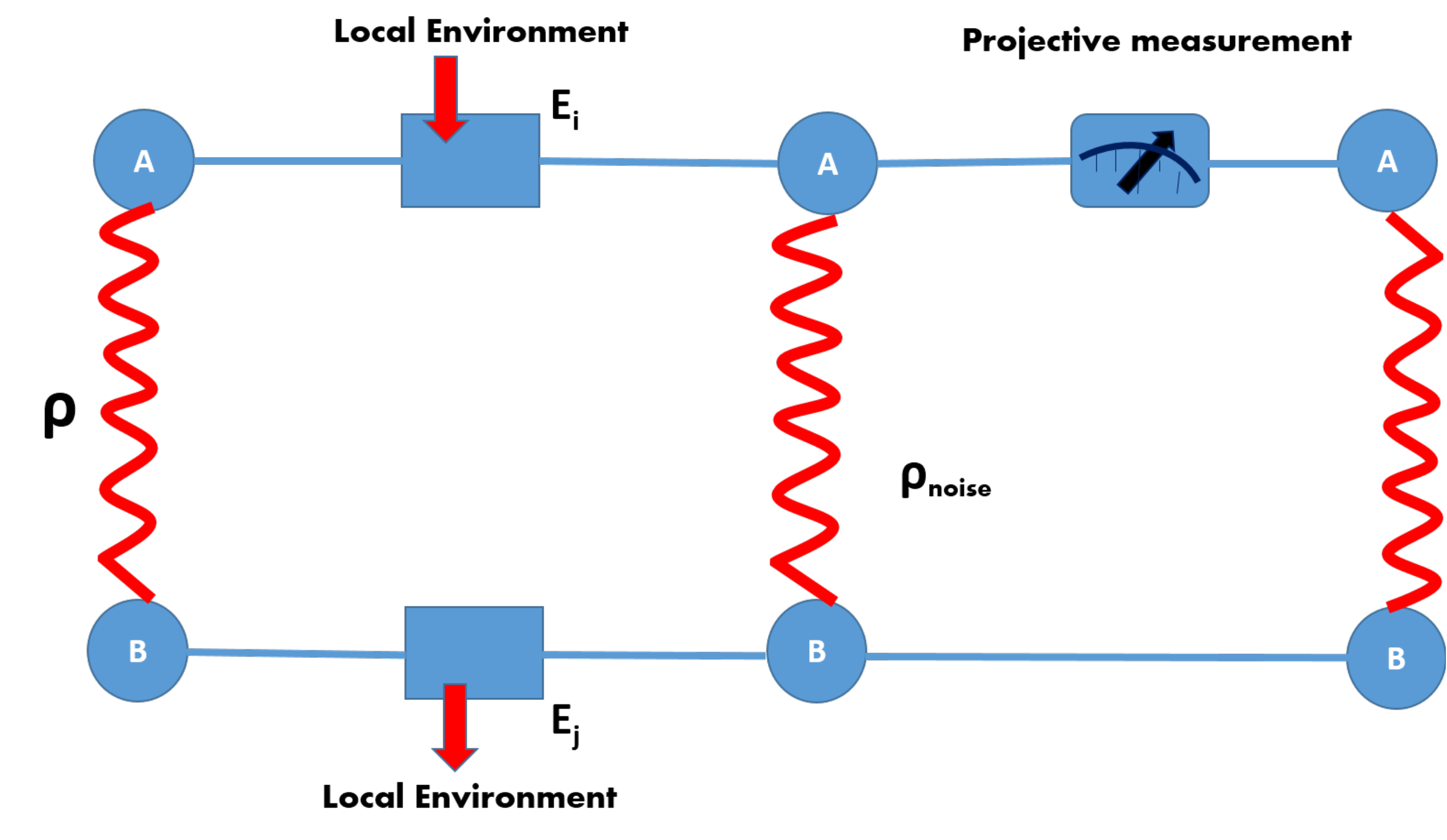}
\caption{(color online)  A two-qubit state interacting with the local environments resulting in an open system (top). The scheme to quantify measurement-induced nonlocality. (bottom).}
\label{open}
\end{figure*}
In general,  an open quantum system refers to a combined model consisting of two quantum bodies, the system $S$ (living in a Hilbert space $\mathcal{H}_S$ ) and the environmnet (living in another Hilbert space $\mathcal{H}_E$ ), such that the entire system $S + E$ lives in the composite space $\mathcal{H} = \mathcal{H}_S \otimes \mathcal{H}_E$. A typical example of the total system $S + E$ is considered isolated, which results in the dynamics of the total system to be unitary. However, the subsystems $S$ and $E$ are not isolated from each other. Due to this unavoidable mutual interaction, the system loses its quantum identities.  In order to understand the effects of environmemt on the dynamical properties of quantum correlation, we consider the Bell diagonal state as an initial state and its Bloch representation can be expressed as 
\begin{equation}
\rho^{BD}=\frac{1}{4}\left[\mathds{1}^a\otimes\mathds{1}^b+\sum^3_{i=1}c_i(0)(\sigma^i \otimes \sigma^i)\right],
\end{equation}
where $\mathds{1}^{a(b)}$ is $2\times 2$ identity operator corresponding to subsystem $a (b)$ and the vector $\vec{c}=(c_1,c_2,c_3) $ is a three dimensional vector composed of correlation coefficients such that $-1\leq c_i(0)=\text{Tr}(\rho^{BD}\sigma^i \otimes\sigma^i)\leq 1$ so that $c_i(0)$ completely specifies the quantum state. Such families of states are indeed nothing but the convex combination of Bell states given as 
\begin{equation}
\rho^{BD}=\lambda^+_{\phi} |\phi^+\rangle \langle \phi^+|+\lambda^-_{\phi} |\phi^-\rangle \langle \phi^-|+\lambda^+_{\psi} |\psi^+\rangle \langle \psi^+|+\lambda^-_{\psi} |\psi^-\rangle \langle \psi^-|,
\end{equation}
where the non-negative eigenvalues of the density matrix $\rho^{BD}$ read 
\begin{equation}
\lambda^{\pm }_{\psi}=[1\pm c_1\mp c_2+c_3]/4, ~~~~~~~~~~ \lambda^{\pm }_{\phi}=[1\pm c_1\pm c_2+c_3]/4 \nonumber
\end{equation}
with $|\psi^{\pm}\rangle=[|00\rangle \pm|11\rangle]/\sqrt{2} $ and $|\phi^{\pm}\rangle=[|01\rangle \pm|10\rangle]/\sqrt{2} $ are the four maximally entangled states. The reduced state  $\rho^{a}=\rho^{b}=\mathds{1}/2$ is a maximally incoherent (mixed) state. In matrix form, we have

\begin{equation}
\rho^{BD}=\frac{1}{4}
\begin{pmatrix}
1+c_3 & 0 & 0 & c_1-c_2 \\
0 & 1-c_3 & c_1+c_2 & 0  \\
0 & c_1+c_2 & 1-c_3 &0  \\
c_1-c_2 & 0 & 0 & 1+c_3
\end{pmatrix}.
\end{equation}

The coupling between the  above  quantum  state  and  environment  can  be  conveniently investigated using Master equation approach. The dynamics of the state can now be written as 
\begin{equation}
\frac{d\rho}{dt}=-\frac{i}{\hbar}[H, \rho]+ \mathcal{L}_{\rho}[t]
\end{equation}
where $\mathcal{L}_{\rho}[t]$ is Lindblad operator which describes the interaction between the state and environment. By choosing appropriate operators corresponding to the environment known as Kraus operators, one can show the equivalence between Master equation approach and Kraus operator-sum representation. In view of the above equivalence, the time evolution of the state is described as \cite{Nielsen2010}
\begin{equation}
\rho(t)=\sum_{i,j}(E_i\otimes E_j)\rho(0)(E_i\otimes E_j)^{\dagger}
\end{equation}
where $\{ E_i\}$ denotes the local Kraus operators characterizing the  noisy  channels  and  it  satisfies  the completely positive and  trace-preserving  map $\sum_iE_i^{\dagger}E_i=\mathds{1}^a$. Proceeding further, we study the influence of bit-phase flip channel, generalized amplitude damping (GAD) and depolarizing noises. Under these noisy environments, it is observed that Bell diagonal state preserves its structure. Time evolved state is then given by
\begin{equation}
\rho^{BD}(t)=\frac{1}{4}
\begin{pmatrix}
1+c_3(t) & 0 & 0 & c_1(t)-c_2(t) \\
0 & 1-c_3(t) & c_1(t)+c_2(t) & 0  \\
0 & c_1(t)+c_2(t) & 1-c_3(t) &0  \\
c_1(t)-c_2(t) & 0 & 0 & 1+c_3(t)
\end{pmatrix} \nonumber
\end{equation}
with the correlation vector $\vec{c}(t)=(c_1(t), c_2(t), c_3(t))$. The concurrence  of Bell diagonal state is computed as 
\begin{equation}
C(\rho^{BD})=2 ~\text{max} \{ 0, |c_1(t)-c_2(t)|  -(1-c_3(t)), | c_1(t)+c_2(t)|  -(1+c_3(t)) \}. \label{Conc}  \nonumber 
\end{equation}
The different MINs are
\begin{eqnarray}
N(\rho)=\frac{1}{4}\left(c_i^2 (t)-c_0^2(t) \right), ~~~~~~~~ 
N_1(\rho^{BD})=c_0(t),      \nonumber \\ 
\text{and}  ~~~~~~~N_{RE}(\rho^{BD})=f(c_0(t))-f(c_1(t),c_2(t),c_3(t))   \nonumber
\end{eqnarray}
where $c_0(t)=\text{min}\{ \lvert c_1(t) \rvert, \lvert c_2(t) \rvert, \lvert c_3(t) \rvert \} $ and 
\begin{equation}
f(c_1(t),c_2(t),c_3(t))=-\left(1+\lambda^{-}_{\psi}\text{log}\lambda^{-}_{\psi}+\lambda^{+}_{\psi}\text{log}\lambda^{+}_{\phi}+\lambda^{-}_{\phi}\text{log}\lambda^{-}_{\phi}+\lambda^{+}_{\phi}\text{log}\lambda^{+}_{\phi}\right) \nonumber
\end{equation}

\begin{figure*}[!ht]
\centering\includegraphics[width=0.7\linewidth]{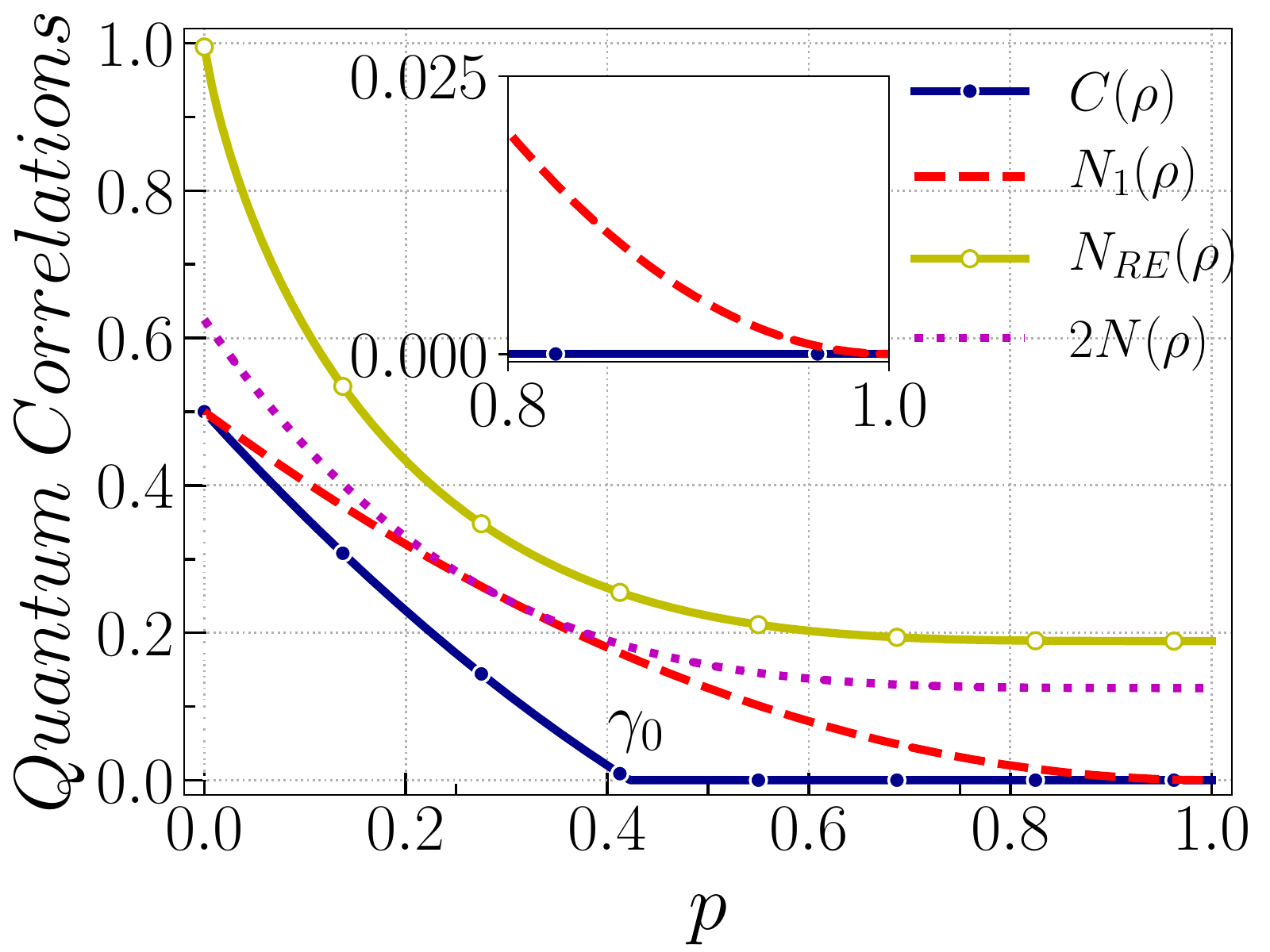}
\caption{(color online) Dynamics of Entanglement and different forms of MIN under Bit-phase flip channel  for mixed partially entangled state with $\vec{c}=(1,0.3,-0.3)$.}
\label{BP}
\end{figure*}

To understand the dynamical behaviors of entanglement and MIN, we consider different noisy channels such as bit-phase flip, depolarizing and generalized amplitude damping channels by considering different initial states. 
 
\textit{\bf Bit--Phase flip channel}: The operation of this channel is described as 
\begin{equation}
\sigma_y|0\rangle \rightarrow \mathrm{i}|1\rangle. \nonumber
\end{equation}
This channel flips the spin and phase of the qubit, with probability $p$ and remains unchanged with probability $1-p$ and the evolved state is written as 
\begin{align} 
\mathcal{E}(\rho)=(1-p)\rho +p ~\sigma_y\rho\sigma_y. \nonumber
\end{align}
Here, we consider the mixed state with coordinates $(1,0.3,-0.3)$ and all the quantities are plotted as a function of decoherence parameter $p$. It is observed that the entanglement and other MIN measures decrease with increase of $p$. The entanglement decreases with increase of $p$. This quantity reaches zero at $p=0.42$ and any further increase of $p$ does not alter the entanglement between the subsystems $a$ and $b$ which is illustrated in Fig. (\ref{BP}). Different forms of MIN also decrease with $p$ and vanish in the asymptotic limit only. This observation shows the presence of quantum correlation beyond entanglement  in the absence of concurrence. 

\textit{\bf Depolarizing channel}:
The depolarizing channel is a noisy channel decohering qubit. This channel is a type of quantum noise which transforms a single qubit into a maximally mixed state 1/2 with probability $\gamma$ and  the qubit remains intact with probability $1-\gamma$ and the channel described by Kraus operators of the following form
\begin{eqnarray}
E_0=\sqrt{1-\gamma}\begin{pmatrix}
    1  &  0      \\
    0  &  1      
\end{pmatrix},
~ 
E_1=\sqrt{\frac{\gamma}{3}}\begin{pmatrix}
    0  &  1     \\
   1 &  0      
\end{pmatrix},\nonumber \\
E_2=\sqrt{\frac{\gamma}{3}}\begin{pmatrix}
    0  &  - \mathrm{i}      \\ 
    \mathrm{i}  &  0      
\end{pmatrix},
~
E_3=\sqrt{\frac{\gamma}{3}}\begin{pmatrix}
    1  &  0      \\
    0 & -1      
\end{pmatrix}.
\end{eqnarray}
where $\gamma=1-\mathrm{e}^{-\gamma't}$ and $\gamma'$ being the damping constant.   Under this noisy channel, the time evolved state coefficients are 
\begin{equation}
c_i(t)=\left(\frac{4\gamma}{3}-1\right)^2 c_i(0) ~~~~~~~~~\text{with}~~ i=1,2,3. \label{destate}
\end{equation}
In order to understand the dynamical behavior of the entanglement (concurrence) and MIN (measured using trace distance, Hilbert-Schmidt norm and relative entropy) under depolarizing noisy channel, we consider different initial conditions $(c_1(0),c_2(0),c_3(0))$. To start with, we consider the initial state with coordinates  $c_1=c_2=c_3=0$, which is a maximally mixed state $\mathds{1}/4$. From eq. (\ref{destate}), it is observed that all the coefficients of evolved state are zero and the time evolved state is also a maximally mixed state. Hence, the entanglement and all forms of MIN remain zero. 

\begin{figure*}[!ht]
\centering\includegraphics[width=0.5\linewidth]{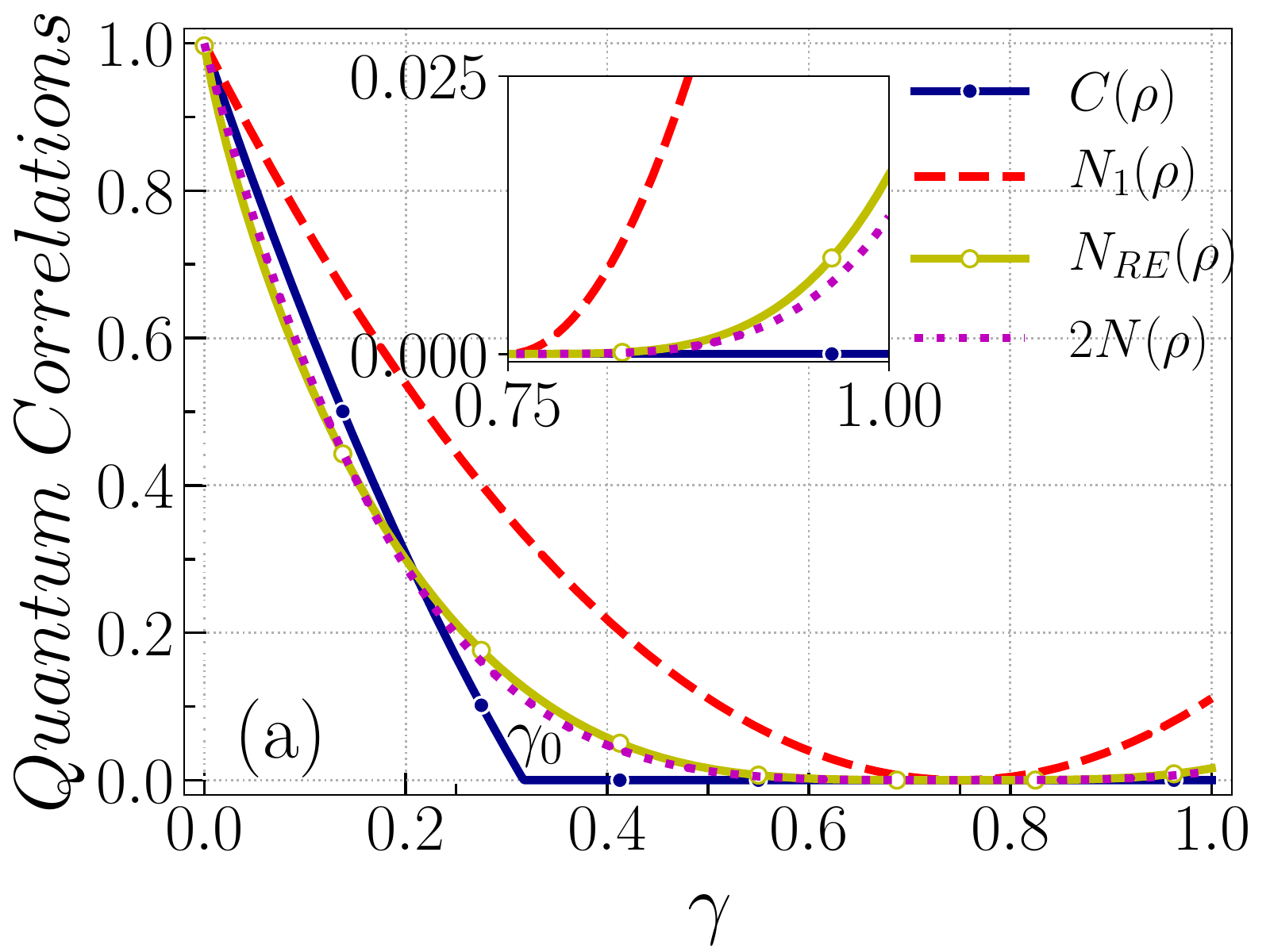}
\centering\includegraphics[width=0.5\linewidth]{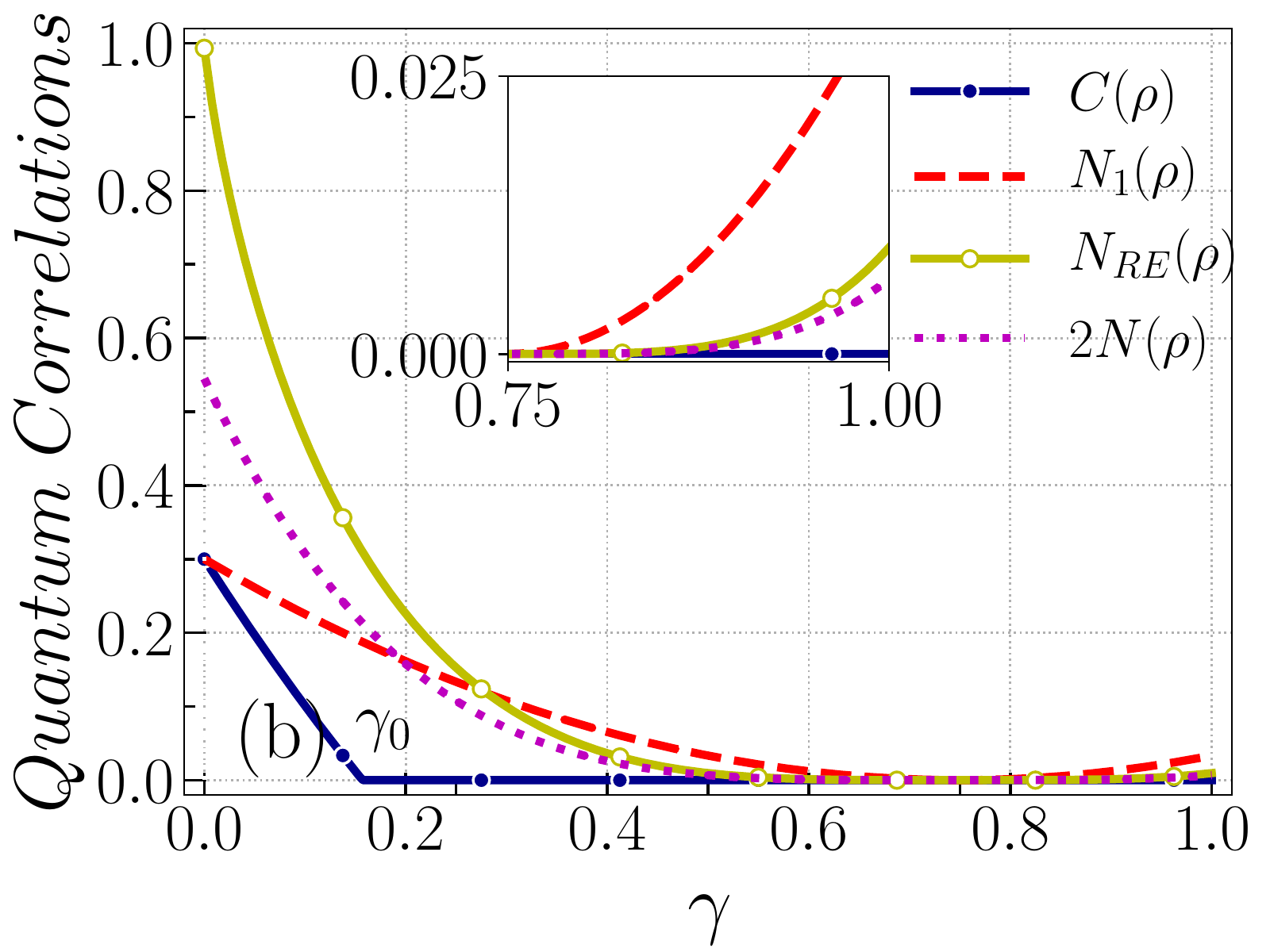}
\centering\includegraphics[width=0.5\linewidth]{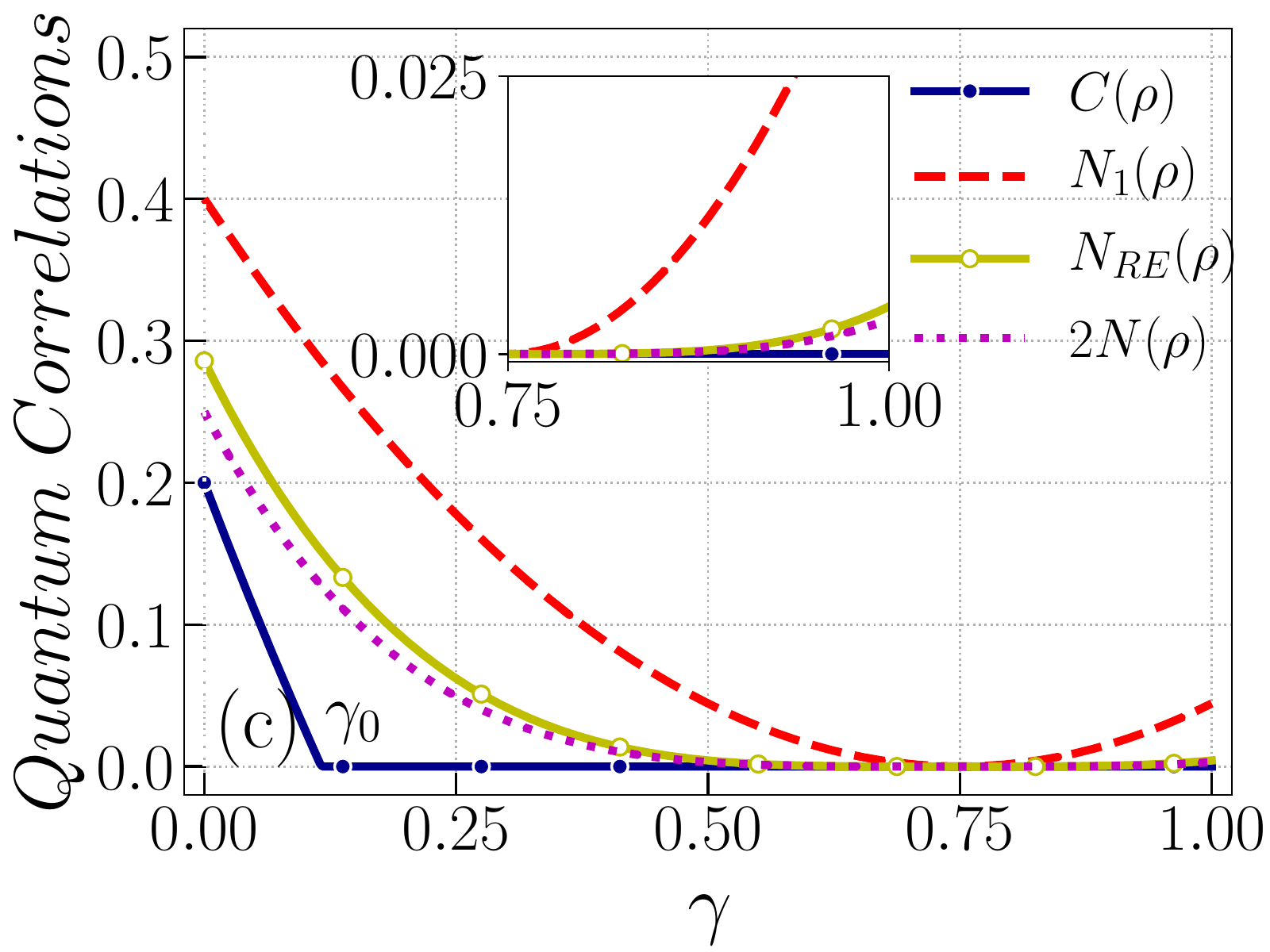}
\caption{(color online) Dynamics of entanglement and different forms of MIN under depolarizing channel for the initial state (a) pure maximally entangled state with  $\vec{c}=(1,1,-1)$, (b) mixed partially entangled states with $\vec{c}=(1,0.3,-0.3)$,  and (c) $\vec{c}=(0.5,-0.4,0.5)$.}
\label{depol}
\end{figure*}

Next, we consider the initial state with the  coordinates assuming any one of the following configuration $(1, 1, -1)$, $(-1, -1, -1)$, $(1, -1, 1)$ and $(-1, 1, 1)$ i.e., $|c_i(0)|=1$, which are pure and maximally entangled states. The concurrence is maximum when $\gamma=0 (t=0)$ and decreases  with increasing of decoherence parameter monotonically. This dynamics of entanglement follows directly from the fact that  $\left(\frac{4\gamma}{3}-1\right)^2$ is a decreasing function of $\gamma$. The concurrence  vanishes at $\gamma\geq \gamma_0$, which is shown by blue line in Fig.(\ref{depol})a. This is known as sudden death of entanglement. In addition, we also observe that the other the MIN quantities also decrease from the maximum monotonically with decoherence parameter and vanishes in a small region of parametric space. Interestingly, all forms of MIN show signs of revival after the dark point while this does not happen in entanglement as shown in inset of Fig.(\ref{depol})a. This implies the fact that the absence of entanglement does not necessarily indicate the absence of nonlocality and show the existence of other quantities.

To enhance our understanding, we now consider partially entangled  mixed states as initial states and we plot the concurrence and MIN as a function of decoherence parameter $\gamma$ as shown in Fig.(\ref{depol})b \& c. At $t=0$, the maximal value of concurrence is less than $1$ due to reduced correlation between the
subsystems. All the MIN quantities decrease with  the decoherence parameter $\gamma$. Here again, the noise cause sudden death in dynamics of entanglement and revival in MIN dynamics.

\textit{\bf Generalized Amplitude Damping}

Next, we consider the generalized amplitude damping (GAD), which models the loss of energy from quantum system to environment at a finite temperature $T$ such as thermal bath. Such a process is described by the Kraus operators \cite{Nielsen2010}
\begin{eqnarray}
E_{0}&=&\sqrt{p}
\begin{pmatrix}
1 & 0\\
0 & \sqrt{1-\gamma }
\end{pmatrix},
~E_{1}=\sqrt{p}
\begin{pmatrix}
0 & \sqrt{\gamma }\\
0 & 0
\end{pmatrix}, \nonumber \\
E_{2}&=&\sqrt{1-p}
\begin{pmatrix}
\sqrt{1-\gamma} & 0\\
0 &  1
\end{pmatrix},
~E_{3}=\sqrt{1-p}
\begin{pmatrix}
0 & 0\\
 \sqrt{\gamma } & 0
\end{pmatrix}, \nonumber
\end{eqnarray} 
where $\gamma =1-\mathrm{e}^{-\gamma' t}$, $\gamma' $ is decay rate and $p$ defines the final probability distribution of stationary state. For simplicity, we fix $p=1/2$ and the  time evolved coefficients are 
\begin{eqnarray}
c'_1(t)=(1-\gamma)c_1(0), ~~c'_2(t)=(1-\gamma)c_2(0),~~\text{and}~~~c'_3(t)=(1-\gamma)^2c_3(0).
\label{GADco} 
\end{eqnarray}

To examine the dynamics of quantum correlation under this channel,  we consider the maximally mixed state $\mathds{1}/4$ as an initial state for our dynamics with the coordinates $c_i(0)=0$. Under this channel again, all the correlation measures become zero (as it is obvious from Eq. (\ref{GADco})).
\begin{figure*}[!ht]
\centering\includegraphics[width=0.5\linewidth]{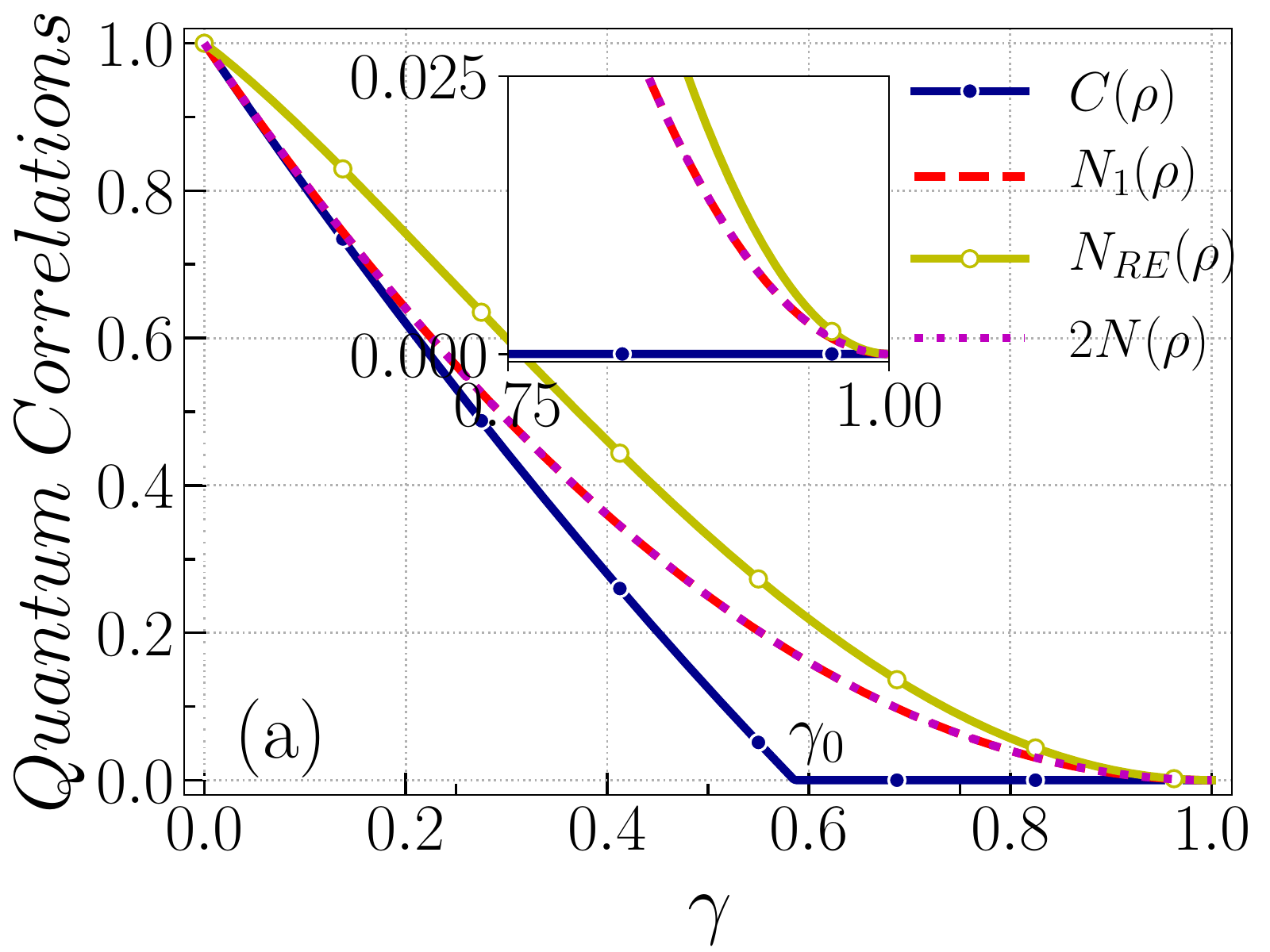}
\centering\includegraphics[width=0.5\linewidth]{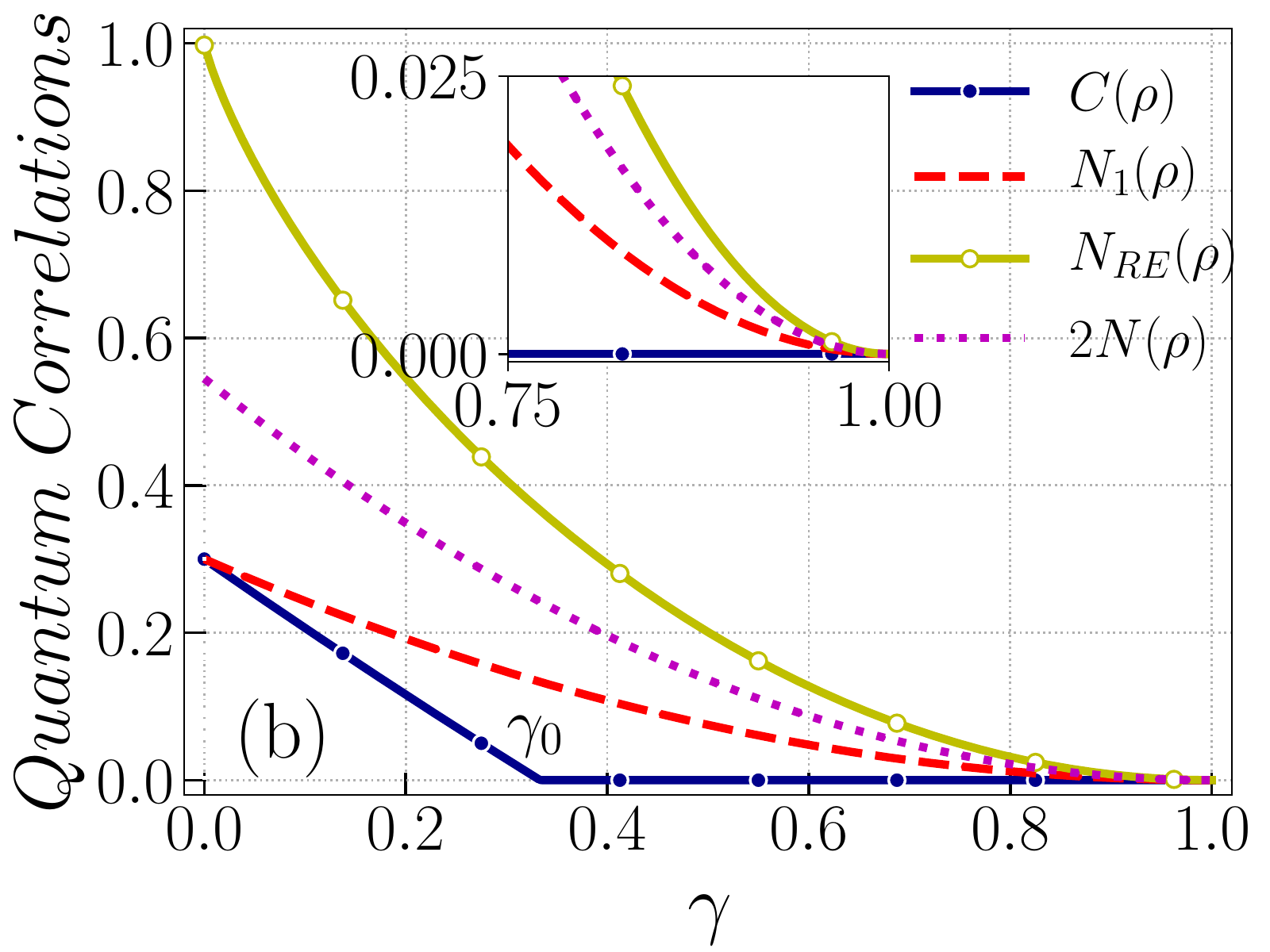}
\centering\includegraphics[width=0.5\linewidth]{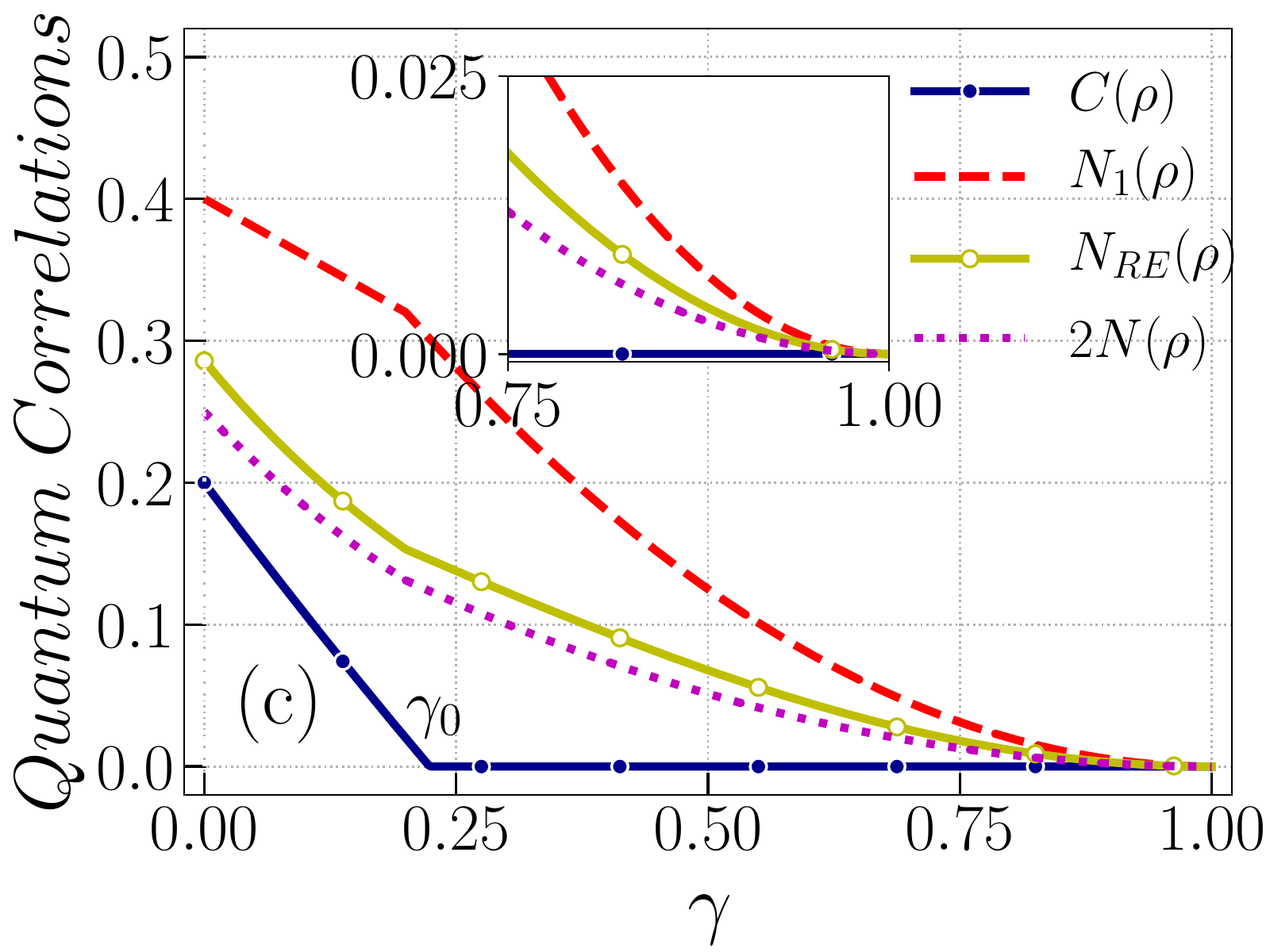}
\caption{(color online) Dynamics of entanglement and different forms of MIN under depolarizing channel for the initial state (a) pure maximally entangled state  with $\vec{c}=(1,1,-1)$, (b) mixed partially entangled states with $\vec{c}=(1,0.3,-0.3)$,  and (c) $\vec{c}=(0.5,-0.4,0.5)$.}
\label{GAD}
\end{figure*}%

Next, we consider pure and maximally entangled states as initial states with the coordinates  being $(1, 1, -1)$, $(-1, -1, -1)$, $(1, -1, 1)$ and $(-1, 1, 1)$. At $t=0 ~(\gamma=0)$, all the correlation measures are maximum. The entanglement (measured by concurrence) decreases with $\gamma$ from maximum and vanishes at $\gamma_0$. This reinforces the fact that the GAD channel also causes sudden death in entanglement similar to bit-phase flip and depolarizing channels. The MIN quantities also decrease with decoherence and vanishes asymptotically as shown in Fig.(\ref{GAD})a. It is observed that the MINs are more robust than the entanglement under GAD channel. 

Next, we consider the mixed partially entangled state as an initial state for our investigation. Except relative entropic MIN (shown in FIg. (\ref{GAD})b), the concurrence and MINs are not maximum in this case due to the mixedness and decrease with decoherence parameter (shown in Fig.(\ref{GAD}) b \& c). Here again, the channel causes sudden death of entanglement, while the other MIN quantities vanish in the asymptotic limit.

\section{Conclusions}
\label{Concl}
In conclusion, we have studied  the dynamics of quantum correlations captured by entanglement and measurement induced nonlocality based on trace distance, Hilbert-Schmidt norm and relative entropy for a two-qubit system coupled to independent Markovian environments. We observe that the noisy channels cause sudden death in entanglement, while MIN quantities are more robust than entanglement. This indicates the presence of nonlocality (in terms of MIN) even in the absence of entanglement between the subsystems. This inherent robustness of MIN can be traced back as a consequence of the negligibility of the set of classical states (zero discord states) in comparison with the whole state space. Our investigations emphasize that an efficient quantum algorithm and subsequent information processing based on the measurement-induced nonlocality offer more resistance to external perturbation and are completely different from that of entanglement. The above observation underscores the robustness of MIN quantities.

%
%

\begin{acknowledgements}
This work has been financially supported by the CSIR EMR Grant No.  03(1456)/19/EMR-II
\end{acknowledgements}



\end{document}